\documentstyle[12pt,epsf,epsfig,wrapfig]{article}
\begin{document}
\phantom{xxxxxxx}

\sloppy
\vspace{5mm}
\begin{flushright}
FUB-HEP/96-15
\end{flushright}
\vspace{2.cm}
\begin{center}
{\large \bf 
Spin-effects and dissociation of space-like photons
\footnote{\normalsize Talk presented by 
Meng Ta-chung
at the 12th {\it International Symposium on High Energy
Spin Physics}, SPIN96, Amsterdam, September 10-14, 1996.}
\footnote{\normalsize Supported in part by Deutsche Forschungsgemeinschaft
(DFG:Me 470/7-2).}
}

 \vspace{5mm}
{\small\it
Institut f\"ur theoretische Physik, FU Berlin,\\ 
Arnimallee 14, 14195 Berlin, Germany
\\ }
\end{center}

\newpage

 By using the characteristic spin-effects observed in 
high-energy hadron-hadron collisions as indicators, hadronic 
dissociation of space-like photons can be directly probed 
by performing measurements in the fragmentation region of 
transversely polarized and unpolarized proton beams, for example,  at
HERA.  This is what my collaborators, and myself 
proposed in a recent paper [1]. This is also what I would like to 
report in this talk.

It is known already for a long time that
hadronic dissociation of space-like
photons may play a
significant role in deep-inelastic
lepton-hadron scattering --- especially
in diffractive processes [2,3].
People seem to agree that,
viewed from the hadron- or nucleus-target,
not only real, but also space-like photons
$(Q^2\equiv -q^2>0$,
where $q$ is the four-momentum of such a photon)
may dissociate into hadronic states. 
But, as far as the following  question is concerned,
different theoretical models (see e.g. Refs. 2-5 and the papers cited
therein)  
seem to give different answers.
How do such hadronic dissociation processes
depend on the standard kinematic
variables of deep-inelastic
lepton-nucleon scattering,  $Q^2$ and
 $x_B\equiv -q^2/(2pq)$,
where $p$ is the four-momentum of
the struck nucleon? Can virtual photons always be treated as 
hadrons in the small $x_B$ and large $Q^2$ region ? 

Viewed from the rest frame of the struck nucleon
mentioned above, the lifetime $\tau_\gamma$ of the virtual hadronic
system 
is of the order $2\nu/Q^2 =1/(Mx_B$), where $\nu$ is the
photon-energy and $M$ is the proton-mass.
This means, the corresponding 
formation/coherence length 
can be much larger than the proton's radius.  
Furthermore, we note that the above-mentioned lifetime 
 $\tau_\gamma$ is a function of $x_B$,
independent of $Q^2$.
Does it imply that
the hadronic dissociation
of a photon always takes place --- independent of
its virtuality $Q^2$? 
Some of the dynamical models
based on such a photon-dissociation picture
(See e.g.[4])  
have been used to describe 
the proton structure function
$F_2^p(x_B,Q^2)$ in the small $x_B$ region
and the obtained results are in reasonable agreement
with the existing data [6,7].
Can we, on the basis of this agreement, say:
``Experiments show that the {\it interaction} between the proton 
and the space-like photon is always {\it hadronic} 
--- independent of the virtuality ($Q^2$) of the latter?''  
Is it correct and/or appropriate to say that, 
in such reactions, the question whether 
we are dealing with {\it hadronic} interactions    depends  on the choice 
of reference frames?  

A large number of inelastic lepton-nucleus 
experiments have been performed in which  
the hadronic properties of the space-like photons have been studied. 
But, probably due to the complicated nuclear structure,  
 the data can be reproduced by different 
models based on different physical pictures [4,5]. 
Can hadronic dissociation of space-like photons be probed without using
nuclear targets ? 

Having the present and the future experimental possibilities at HERA  
in mind, we think it would be useful to 
consider the characteristic spin-effects observed --- and only 
observed --- in the fragmentation regions of hadron-hadron collisions 
at comparable energies and use them as indicators to probe the   
space-like photons at given values of $Q^2$ and $x_B$. 
To be more precise, we propose to measure  the left-right asymmetry 
$A_N$ of produced charged mesons  in the fragmentation region
of the transversely polarized proton $p(\uparrow )$, 
 to measure the $\Lambda$-polarization $P_{\Lambda}$ 
 of the unpolarized protons $p$ at HERA
in the small-$x_B$ region for different $Q^2$-values, and
to compare the obtained
results with those obtained
in the corresponding hadron-hadron collisions.

In order to demonstrate in a quantitative manner
how the $Q^2$-dependence of such dissociation
processes may manifest itself, we
examine the $F_2^p(x_B,Q^2)$-data [6,7] in the
small-$x_B$ region.
In Fig.1, we separate the well-known vector-dominance
contribution (See e.g. [4,5])
from ``the rest,''
and we consider the following two extreme possibilities which correspond to
two very much different physical pictures:
(i) The hadronic dissociation of virtual space-like
$(Q^2>0)$ photons take place
for all possible $Q^2$-values. 
In other words, in this picture $\gamma^*(Q^2)$ should
always be considered as a hadronic system --- independent of $Q^2$.
(ii) The hadronic dissociation 
of such photons depends very much on $Q^2$.
In terms of a two-component picture (See e.g.[2-5])
 the virtual photon $\gamma^*(Q^2)$
 is considered to be either in the
\begin{wrapfigure}{r}{7cm}
\epsfig{figure=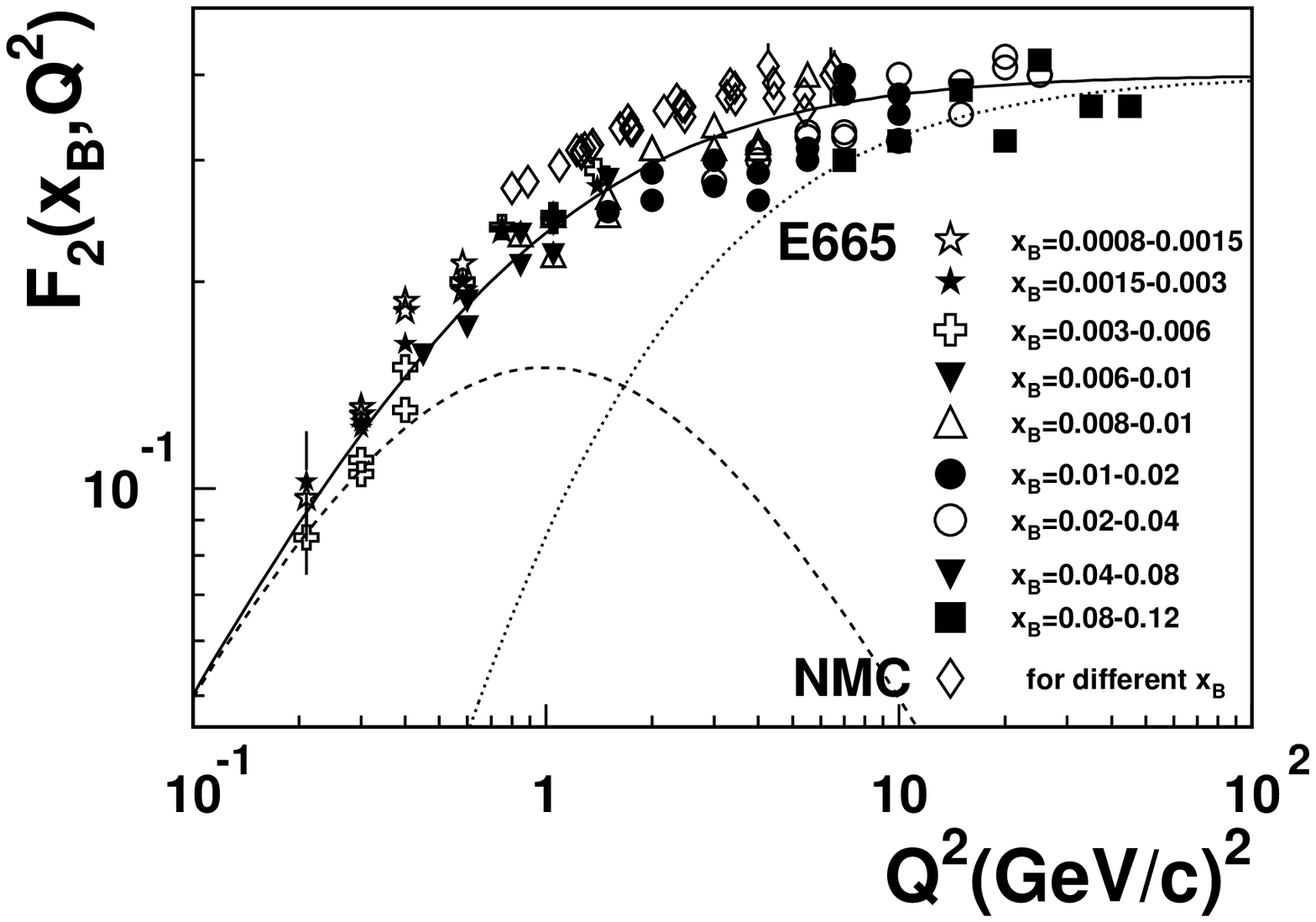,width=7cm}
{\small Figure 1: Structure function $F_2^p(x_B,Q^2)$ as a function of $Q^2$.
The data-points are taken from [6,7]; and they are parametrized
(shown as solid line) in order to carry out the quantitative calculation
mentioned in the text.
The dashed line is the contribution from the
vector meson dominance. The difference, which is called ``the rest'', is
shown as dotted line.  }
\end{wrapfigure}
 ``bare photon'' state or
 in a hadronically dissociated state (``hadronic cloud'') described
 by the vector-dominance
 model.  
In other words,  only the latter can be considered hadronic; and the
probability 
for $\gamma^*(Q^2)$to be in this state is the ratio between the values 
shown by the dashed and the full lines.

Let us first look at the
left-right asymmetry data [8] for
$\pi^{\pm}$-production in
$p(\uparrow )+p$ and see what we may obtain by
replacing the unpolarized proton-target $p$ by
a photon with given $Q^2$, $\gamma ^*(Q^2)$.
It is clear that the corresponding asymmetry which we denote by
$A_N(x_F|Q^2)$ will have the following properties:
If scenario (i) is correct, we shall see no
change in $A_N(x_F|Q^2)$ by varying $Q^2$.
If scenario (ii) is true, there will be a 
significant $Q^2$-dependence. 
The results are shown in Fig.2a.
Similar effects are expected also for $K^+$-mesons. 
In order to emphasize the model-independence of this test,  
 the curve which goes through 
the existing proton-proton data points [8] should be considered 
as an empirical fit, although such an asymmetry-data can be described 
by a relativistic  quark model [9].

We next consider the $\Lambda$-polarization
$P_\Lambda(x_F,Q^2)$
in the process $p+\gamma^*(Q^2)\to \Lambda +X$
in which unpolarized proton beam is used [10].
Also here, we expect to see {\it no $Q^2$-dependence
for scenario (i) but a significant $Q^2$-dependence
for scenario (ii). }
This is shown in Fig.2b.

\begin{tabular}{cc}
\begin{minipage}{6.cm}
\psfig{file=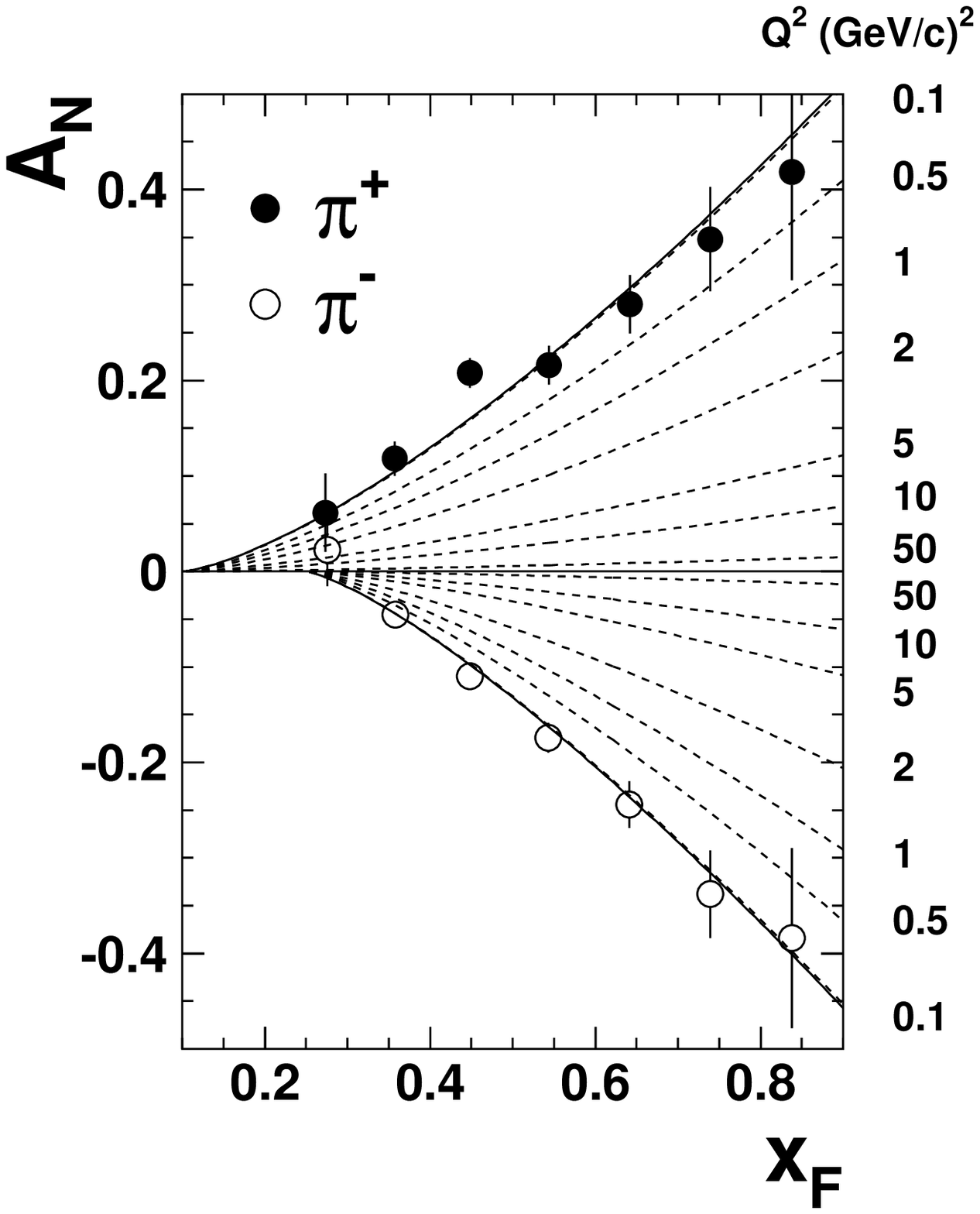,width=6cm}
\end{minipage}
&
\begin{minipage}{6.cm}
\psfig{file=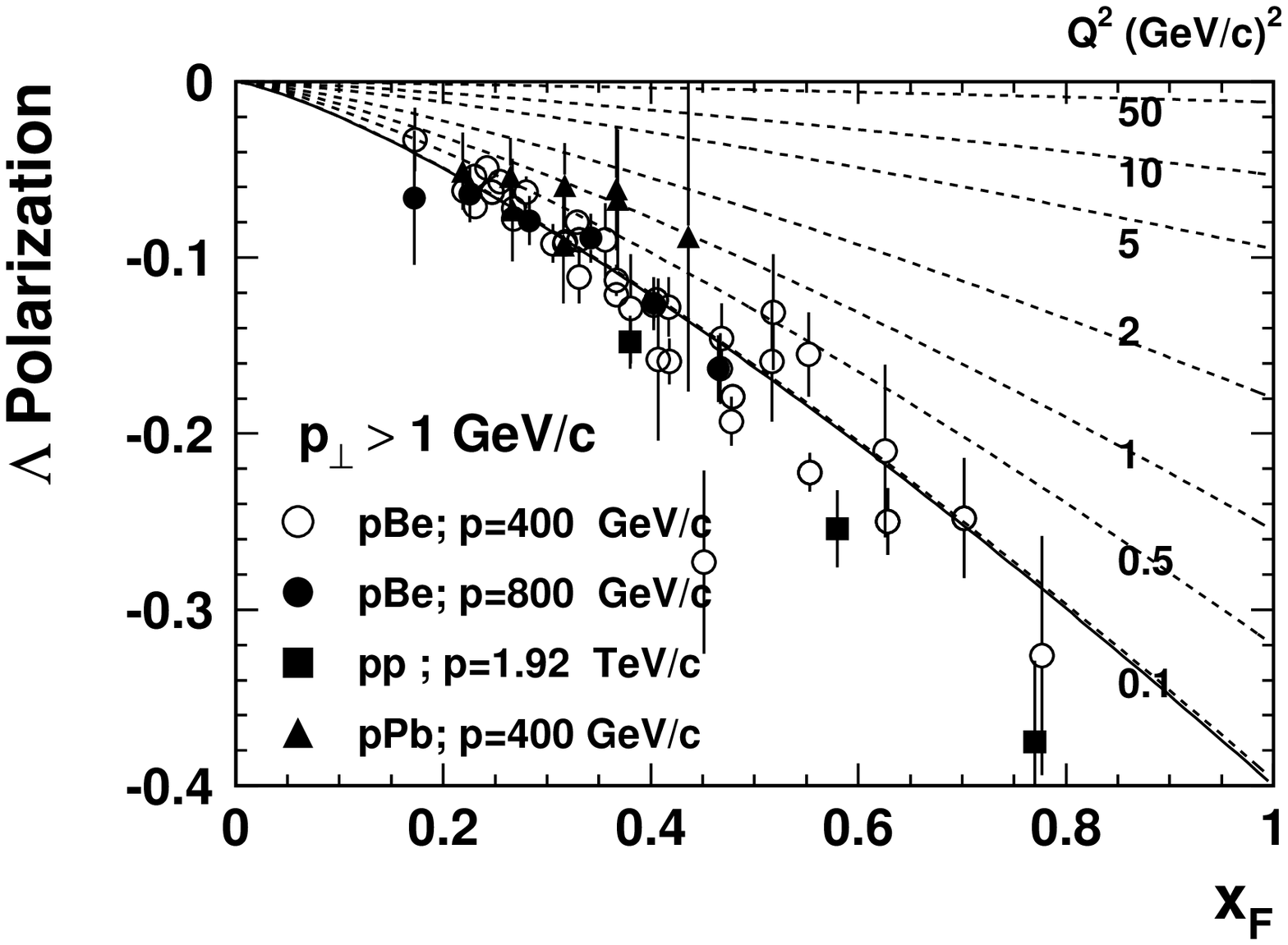,width=8cm}
\end{minipage}
\end{tabular}

\noindent  {\small Figure.2: (a)
Left-right asymmetry for pion-production in
$p(\uparrow ) +\gamma^*  \rightarrow \pi^\pm +X$ as function of
$x_F$ at different values of $Q^2$. The data are for $p(\uparrow ) +p
\rightarrow \pi^\pm +X$ and are from [8].}
{\small (b)
Polarization for $\Lambda$-production in
$p +\gamma^*  \rightarrow \Lambda  +X$ as  function of
$x_F$ at different $Q^2$. The data for the indicated proton-proton and 
proton-nucleus collisions are 
from [10].}



\vfill
{\small\begin{description}
\item{[1]} C.~Boros, 
Z.~Liang and T.~Meng, preprint FUB-HEP/95-21 (revised version). 
\item {[2]} H.T. Nieh, Phys. Rev. D1, 3161 (1970);
         Phys. Rev. D7, 3401 (1973). 
\item {[3]} T. Bauer, R. Spital, D. Yennie and F. Pipkin,
         Rev.Mod.Phys. {\bf 50}, 261 (1978).
\item {[4]} See e.g. G. Piller, W. Ratzka and W. Weise,
          Z. f. Phys. {\bf A352}, 427 (1995).
\item {[5]} See e.g. M. Arneodo, Phys. Rep. {\bf 240},
301 (1994). 
\item {[6]} NMC Coll., P. Amaudruz et al., Phys. Lett.
          {\bf 295B}, 3 (1992). 
\item {[7]} Fermilab E665 Coll., A. V. Kotwal,
                       in Proc. of the  Fermilab Conf-94/251-E (1994).
\item {[8]} FNAL E581/704 Collaboration, see e.g. 
    A. Bravar et al., Phys. Rev. Lett.
      {\bf 75}, 3073 (1995) and the papers cited therein. 
\item {[9]} C. Boros, Z. Liang and T. Meng, Phys. Rev. Lett. {\bf 70},
1751 (1993), Phys.Rev.{\bf D51}, 4867 (1995); Phys.Rev.{\bf D} (in
press). 
\item {[10]} K. Heller, in 
  Proc. of this Symposium and the papers cited therein. 

\end{description}
\end{document}